# Regional Economic Convergence and Spatial Quantile Regression

**ABSTRACT**

The presence of *β*-convergence in European regions is an important issue to be analyzed. In this paper, we adopt a quantile regression approach in analyzing economic convergence. While previous work has performed quantile regression at the national level, we focus on 187 European NUTS2 regions for the period 1981-2009 and use spatial quantile regression to account for spatial dependence.

**Authors:**

1 Alfredo Cartone

University of Chieti-Pescara

Department of Economic Studies

alfredo.cartone@unich.it

2 Geoffrey J.D Hewings

University of Illinois at Urbana-Champaign

Regional Economics Applications Laboratory

hewings@illinois.edu

**3** Paolo Postiglione

University of Chieti-Pescara

Department of Economic Studies

Viale Pindaro, 42 - 65127 Pescara, Italy

e-mail: postigli@unich.it

Tel.:  +39 085 4537939

# 1. Introduction

Attention to issues of inequality has become a topic of considerable interest in both developing and developed economies. However, much of the research has focused on interpersonal inequality with explorations of changes in standard measures at the national level within countries. A secondary and equally important dimension is the nature of spatial inequality. Over the last three decades, there have been many significant contributions (Abraham and Von Rompuy 1995; Barca et al. 2012; De Dominicis 2014). These contributions have extended the standard Barro and Sala-i-Martin (1995) approach to account for the presence of spatial externalities; in some cases, authors have sought to address the problem of heterogeneity by adopting notions of club convergence. In this paper, two additional perspectives are introduced that together offer important new insights into the process of regional convergence/divergence in Europe. First, a spatial quantile regression methodology is used to account for variations across the distribution of regions rather than adopting the more restricted analysis based on means.

The paper is organized as follows. Section 2 presents a review of the essential contribution in the literature. Section 3 describes the economic model, while section 4 depicts the spatial quantile regression in the case of a Mankiw-Romer-Weil (1992) (hereafter, MRW) economic convergence model. Section 5 includes application of spatial quantile regression to European NUTS 2 regions to study cross-sectional $\beta$-convergence. Section 6 concludes.

# 2. Quantile Empirics of Economic Convergence

The debate on economic growth and economic convergence has been increasing since the early work of Solow (1956). The hypothesis of economic convergence has been studied both under the lens of exogenous and endogenous growth models. Exogenous growth models assume that growth is driven by exogenous technological progress in which capital accumulation guarantees economic convergence. Conversely, endogenous growth models highlight the importance of human and physical capital, wherein the potential lack of convergence results from increasing returns to capital (Romer 1986).

A wide discussion about the methodological drawbacks of different techniques adopted in the empirics of economic convergence have also characterized the last thirty years. The $\beta$-convergence and the $\sigma$-convergence measures represent the most frequently used metrics to test the presence of economic convergence. $\sigma$-convergence implies a reduction of the standard deviation of output per capita over time while absolute $\beta$-convergence is tested by regressing the growth rate for a cross section of countries (or regions) on the initial levels of the output. Finally, conditional $\beta$-convergence represents a natural extension of the absolute one while controlling for other regressors (Barro et al. 1991).

An alternative approach has its roots in time series econometrics. Bernard and Durlauf (1995), among others, offer an analysis based on the application of a unit root test for which the log of the difference between two economies may differ from zero for a while but is expected to be a zero-mean stationary process in presence of convergence. Hence, this gap should not contain a deterministic or a stochastic trend. Le Pen (2011) applies units roots and stationarity tests introducing structural breaks in the deterministic component and enhancing Pesaran (2007), whose technique may neglect potential changes in the structural parameters of regional economies. This leads to the discovery of higher convergence properties in contrast to previous results and to reduce in magnitude the impact of transitional shocks as time passes.

In order to carry on a convergence, test able to consider temporal transitions heterogeneity and cross-section heterogeneity at the national level, Monfort et al. (2013) apply the Phillips and Sul (2007, 2009) methodology to a panel of European Countries. In fact, as highlighted by Phillips and Sul (2009), traditional tests based on stochastic convergence might not be adequate in the presence of differences in technology and varying speeds of convergence. Results from Monfort et al. (2013) find that convergence is strictly linked to the identification of convergence clubs in the European Union and that there exist significative differences between Eastern and Western Europe.

Not surprisingly in recent years, part of the interest of researchers and policy makers has moved towards potential consequences of spatial interactions and stressed the use of spatial models for modelling regional growth (Fingleton and López-Bazo 2006; Rey and Montouri 1999). In fact, the role of spatial interactions cannot be directly retrieved neither in an analysis based on stochastic convergence nor in a test by non-spatial models of *β*-convergence. Indeed, the spatial analysis of spillover effects remains an important topic especially in the context of highlighting the effect of spatial externalities resulting, for example, from knowledge spillovers (Ertur and Koch 2007).

Another basic assumption of *β*-convergence, that all economies have a common linear specification, has been identified as problematic in the recent literature (see Azomahou et al. 2011). Linear models cannot highlight the potential heterogeneity that appears in a cross section of units and, for this reason, semi-parametric and non-parametric techniques have been applied to verify the presence of economic convergence while tackling the presence of non-linearities and without the restriction of specifying the income distribution and the functional form of the regression. Furthermore, the use of linear regression for testing economic convergence has been challenged by Quah (1993a, b) and Friedman (1992) who argued that the results of linear regression may be affected by *regression fallacies*, particularly referring to the independent and identically distributed (i.e., *iid*) assumption and to the average causal effect assumed by linear regression. Those findings stress the possibility of considering structural differences and heterogeneities as sources of instability and for investigating different responses from the one obtained if we assume that the process could be modeled with a common stationary mean.

Following this direction, quantile regression (Koenker and Basset 1978) represents a suitable tool to analyze the various effects of covariates on growth levels beyond the common average effect (Koenker 2005). This technique adopts a semi-parametric framework in which we can display a more robust coefficient of median global convergence and shape more accurately significant tendencies beyond a conditional mean parameter. Therefore, the key issue of quantile regression is the possibility to analyze some significant tail effects undetectable in linear regression and potential heterogeneous effects hidden in the linear estimation.

The use of quantile regression represents a different perspective to analyze the presence of heterogenous effects in the conditional distribution. In contrast to the use of other non-parametric and semi-parametric techniques that account for non-linearities in the convergence process (Azomahou et al. 2011; Li et al. 2016), quantile regression models the growth curve by different quantiles, identifying heterogenous effects at different levels of the conditional distribution. Hence, quantile regression allows us to explore potential differences in the convergence rate depending on the conditional quantile estimated. Moreover, this is pursued in a heteroscedasticity-robust and distribution-free frame.

Only few studies have exploited a more comprehensive definition of economic growth using quantile regression (Barreto and Hughes 2004; Canarella and Pollard 2004; Cunningham 2003; Dufrenot et al. 2010; Foster 2008; Mello and Perrelli 2003). Additionally, these studies have not considered the presence of spatial interactions in the quantile approach (Kostov and Le Gallo 2015). Therefore, a spatially augmented alternative to standard quantile regression can be adopted by introducing spatial dependence into a quantile framework (McMillen 2013) thereby avoiding the problem of the quantile regression estimator being inconsistent due to endogeneity (Chernozhukov and Hansen 2006; Kim and Muller 2004). In this sense, another issue that can be explored by spatial quantile regression is a greater understanding of whether the spatial autocorrelation changes with the quantile level. This could also highlight further source of differences in the convergence rates depending on the spatial interactions.

Although there have been a large number of studies, economic convergence at the European regional level remains one important question in the debate about policy effectiveness (Boldrin and Canova 2001). Especially in the case of Europe, regional economic convergence is of primary relevance as the Treaty of the European Union states the necessity for reducing differences in economic performance. Therefore, testing the presence of economic convergence at the regional level while accounting for spatial effects would provide important, strategic information for understanding policy implications (Le Gallo and Dall'erba 2006). Thence, we perform a spatial quantile cross-sectional analyses at European NUTS 2 level during the period 1981-2009 to investigate the presence of quantiles differences, avoiding Galton's fallacy, and shedding more light on possible heterogeneous effects of covariates.

Finally, we exploit the main findings from spatial quantile regression analysis to develop an empirical method to identify clusters of regions. This approach enhances the recent literature interested in a deeper analysis of the joint effects of spatial dependence and heterogeneity in the case of the economic growth (Andreano et al. 2017; Postiglione et al. 2013). Beyond a mere identification of spatial groups, analyzing results of spatial quantile regression in terms of clusters allows analysts and policy makers to focus on differences based on different economic motions at different quantile levels.

## 3. The Economic Model

Neoclassical growth theory was introduced by Solow (1956) under the main assumptions of an exogenous saving rate, decreasing productivity of capital, and constant returns to scale and under the hypothesis that poorer countries would catch up with the richest ones. The most popular model to analyze economic growth following this Neoclassical framework is based on the definition of a conditional growth function that allows for different covariates as a reason for the discrepancy between spatial units (Barro and Sala-i-Martin 1995). Conditional convergence claims the possibility of different steady states and is suitable for the case of testing convergence between regions located in different countries.

Mankiw et al. (1992) define conditional growth function as:

$$g_i = \alpha + \beta q_i + \pi_1 v_i + \pi_2 s_i + \pi_3 h_i + \varepsilon_i \qquad (1)$$

where $g_i$ is the average growth rate of GDP per worker for each unit over the period 1981-2009; $q_i$ is the natural logarithm of the initial GDP per worker level; $\beta = -(1 - e^{-\lambda T})$, where $t$ denotes the period and $\lambda$ the speed of convergence; $v_i = \ln(n_i + l_i + d_i)$ includes $n_i$ as the population growth rate,; $l_i$ is the level of technology; $d_i$ is the depreciation rate of capital; $s_i$ is the natural logarithm of saving; $h_i$ is a measure of human capital, with $l_i + d_i = 0.05$ according to Mankiw et al. (1992), and $\varepsilon_i$ is the normal error term. Testing this convergence hypothesis necessitates a search for a negative correlation between the GDP per capita growth rates and the initial level of GDP per capita.

When the model is specified for spatial units, convergence models are likely to be affected from spatial effects such as spatial dependence and spatial heterogeneity (Anselin 1988). Spatially augmented models have been proposed to embed the spatial dependence in the growth function due to spatial proximity and spillovers, model misspecification, and a variety of measurement problems. Furthermore, the presence of spatial heterogeneity in economic growth models has been properly investigated only in recent times (Andreano et al. 2017; Postiglione et al. 2013) with the aim of analyzing structural heterogeneities that can be explained by the presence of multiple regimes in the growth function or with convergence clubs (Durlauf and Johnson 1995).

Several spatially augmented specifications of growth function have been considered in the literature. A well-known spatial specification refers to the SAR model (LeSage and Pace

2009) that can be expressed in matrix notation, in the case of conditional $\beta$-convergence, as follows:

$$\mathbf{g} = \rho \mathbf{W}\mathbf{g} + \mathbf{X}\boldsymbol{\theta} + \boldsymbol{\epsilon} \qquad (2)$$

where $\rho$ is the spatial autoregressive parameter, $\mathbf{W}$ is the $n \times n$ weight spatial matrix that defines the connectivity between spatial units, $\boldsymbol{\theta}$ is the vector of parameters corresponding to the model covariates, $\mathbf{X}$ is the $n \times (p+1)$ matrix with $p$ explanatory variables (in our case $p = 4$), and $\boldsymbol{\epsilon}$ is the usual normal error term.

In this paper, our aim is to test for conditional convergence at different quantile levels by considering spatial dependence. The SAR model represents the basic model for our estimation of spatial quantile regression model. This argument will be elaborated in the next section.

## 4. Quantile Regression for Spatial Data and Identification of Spatial Clusters

In the economic growth literature, linear models have played a key role in testing for the presence of convergence tendencies (Barro and Sala-i-Martin 1995). However, the use of linear models raises several problems, including the well-known Galton's fallacy (Friedman 1992; Quah 1993a). The main problem is that linear models offer only a partial view of the phenomenon. In the case of a $\beta$-convergence model, the linear model only defines the effects of the covariates on the expected value of the GDP per worker growth rate as:

$$E(\mathbf{g}|\mathbf{X}) = \mathbf{X}\boldsymbol{\theta} \qquad (3)$$

In this sense, linear models in general can only unveil information about mean effects, while no more is offered to understand the effects at different levels of the distribution while quantile regression can overcome the limits related to the use of linear models (Canarella and Pollard 2004; Cunningham 2003).

Generally speaking, each random variable is characterized by a probability function as:

$$F(g) = \Pr(G < g) \qquad (4)$$

We can derive the quantile $\tau$ of any random variable $G$ from the probability function as:

$$Q_G(\tau) = inf\{g : F(g) \geq \tau\} \qquad (5)$$

According to this way of characterizing a probability function, quantile regression expresses the quantiles of a conditional distribution as (Koenker 2005):

$$Q_G(\tau | \mathbf{x}_i) = \mathbf{x}_i^t \boldsymbol{\theta}_\tau \qquad (6)$$

where $\mathbf{x}_i^t$ is the covariates vector for the $i = 1, \ldots, n$ units. Equation (6) defines the $\tau$-quantile of the $(\mathbf{g}|\mathbf{X})$ conditional distribution, and $\boldsymbol{\theta}_\tau$ is the vector of estimated coefficients corresponding to the covariates selected for our model.

A more familiar way to express equation (6) was proposed by Buchinsky (1998):

$$g_i = \mathbf{x}_i^t \boldsymbol{\theta}_\tau + u_{\tau,i} \qquad (7)$$

where $u_{\tau,i}$ corresponds to the error term on which no further assumptions different from the ones associated with (6) and (7) are made, thus $Q_U(\tau | \mathbf{x}_i) = 0$.

Quantile regression may be estimated when the distribution of the error is known (Geraci and Bottai 2006). However, Koenker and Basset (1978) initially proposed a robust solution to estimate quantile models without any assumption on the distribution of the error term. This solution is obtained by solving the following minimum:

$$\min_{\boldsymbol{\theta}} \frac{1}{n} \left\{ \sum_{i: g_i \geq \mathbf{x}_i^t \boldsymbol{\theta}} \tau |g_i - \mathbf{x}_i^t \boldsymbol{\theta}| + \sum_{i: g_i < \mathbf{x}_i^t \boldsymbol{\theta}} (1 - \tau) |g_i - \mathbf{x}_i^t \boldsymbol{\theta}| \right\} \qquad (8)$$

using linear programming (Koenker and D'Orey 1987).

Even if the standard specification of quantiles model is very useful to understand wider effects at diverse levels of the conditional distribution, the estimation of the growth function involving spatial units may be affected by the presence of spatial dependence (Panzera and Postiglione 2014; Rey and Montouri 1999).

In the case of European economic convergence, in this paper, we propose the use of a spatial autoregressive quantile model augmented for including a spatial lag as defined by McMillen (2013):

$$\mathbf{g} = \rho_\tau \mathbf{W}\mathbf{g} + \mathbf{X}\boldsymbol{\theta}_\tau + \mathbf{u}_\tau \qquad (9)$$

where, in addition to the specification of the non-spatial quantile model, we can note a spatial autoregressive parameter $\rho_\tau$ at the selected quantile level $\tau$ and the weight spatial matrix $\mathbf{W}$. Concerning the error term, the following condition is satisfied:

$$Q_U(\tau|\mathbf{x}, \mathbf{W}\mathbf{g}) = 0 \qquad (10)$$

As noted in Chernozhukov and Hansen (2005), the presence of endogeneity in the form of the spatial lag variable $\mathbf{W}\mathbf{g}$ affects the conventional quantile regression estimator and makes it inconsistent. The impact of endogenous variables has been discussed in the econometrics literature, including the case when endogeneity is embedded in the quantile model by the presence of spatial effects (McMillen 2013).

Several methods can be used to evaluate the impact of endogenous variables and, in our case, of the spatial autocorrelation parameter. Kim and Muller (2004) propose a double stage quantile regression (DSQR) estimation method that can be considered the analogue of the two stages least squares procedure for the spatial linear model defined by Kelejian and Prucha (1998). DSQR estimates parameters using quantile regressions with random regressors after the endogeneity is treated via preliminary predictive quantile regressions. DSQR is used in spatial applications (Zietz et al. 2008) and is particularly attractive in terms of computational efficiency; nevertheless, the asymptotic properties of the estimator assume that the data are *iid*.

Chernozhukov and Hansen (2005) suggest an instrumental variable quantile regression (IVQR) estimator that is a modified version of the standard quantile regression estimator to address the presence of endogenous variables. This estimator does not assume independence among observations and ensures good performance in the case of weak instruments. The IVQR estimator has been shown to be consistent and asymptotically Gaussian under appropriate regularity and identification conditions. Additionally, while inference for the DSQR estimator is based on bootstrap, the IVQR estimator inference is based on the estimation of covariance matrix of estimated parameters on which we rely to derive confidence intervals and inference (Chernozhukov and Hansen 2006; McMillen 2013).

A different approach to spatial quantile regression estimation is offered by Kostov (2013) who introduces an empirical likelihood quantile regression (ELQR) estimator, a non-parametric analogue of likelihood estimation for the linear model. If the ELQR estimator remains a valid alternative in the case of multiple endogenous variables, this estimator produces greater computational costs, and may not be preferable with regards to the IVQR method, which has similar properties.

Therefore, to avoid the consequences of a weak specification of the instruments, the IVQR approach is preferred in this paper, considering its properties and computational costs. Finally, for the spatial specification under investigation, the matrix of instruments **Z** is suggested by McMillen (2013) as:

$$\mathbf{Z} = [\mathbf{X}, \mathbf{WX}, \mathbf{W}^2\mathbf{X}] \qquad (11)$$

like those defined by Kelejian and Prucha (1999) for the spatial linear model.

However, the spatial dependence is not the only effect that influences geographically distributed data. It is important to also consider the spatial heterogeneity. In the regional convergence analysis, the existence of structural heterogeneities may be explained, for example, with convergence clubs or cluster of regions that share the same growth path (Postiglione et al. 2013).

Further, a cluster of regions may be composed of units that are not necessarily contiguous and they may not belong to the same country. This matter is of primary relevance in the European Union for the definition of appropriate supranational policy and for the regulations for the provision of funds to the regions of the EU member countries. Section 5 provides further discussion about this topic.

In this paper, to identify clusters of regions, we propose an empirical approach that is derived from the study of efficiency in a frontier production framework (Tauer 2016). Consider one of the upper quantiles, $\tau_u$, chosen to represent a higher level of conditional distribution of $g_i$ (i.e., of the economic growth). The regions are classified in diverse groups according different interval of residuals from linear quantile estimation at the opted level. In fact, given a set of characteristics addressed in the model by a certain level of the covariates, regions in the higher part of the conditional distribution may be considered as better performing compared to the others.

## 5. Empirical Results

Data for 187 NUTS 2 European regions are collected for 12 countries (i.e., Austria, Belgium, Finland, France, Western Germany, Greece, Italy, Portugal, Spain, Sweden, the Netherlands, and the United Kingdom) with the goal of verifying the presence of conditional $\beta$-convergence over the period 1981-2009. The spatial scale adopted is the NUTS2 level as it is considered the most appropriate for regional convergence studies (Le Gallo and Dall'erba 2006). In fact, NUTS2 units represent a relevant level for consideration of policies for development funding and subsidies (i.e., assignment of structural funds of European Commission).

The model includes the log of the rate of growth of the GDP per worker ($q$) as the dependent variable and the set of covariates is composed by saving rates ($s$), the rate of population growth ($n$), the technological progress rate ($l$), the capital depreciation ($d$), and the natural logarithm of the human capital ($h$).

Results for quantile regression conditional convergence in European regions represent a scenario that expands the one obtained through linear regression. This feature generates additional information of the effects at different quantile levels. Estimation of quantile regression model is obtained by using the methodology described in Koenker and d'Orey (1987).

In figure 1, the estimated quantile regression effects for MRW model are described. Confidence intervals of the parameters are obtained by inverting the rank test (Koenker 1994).

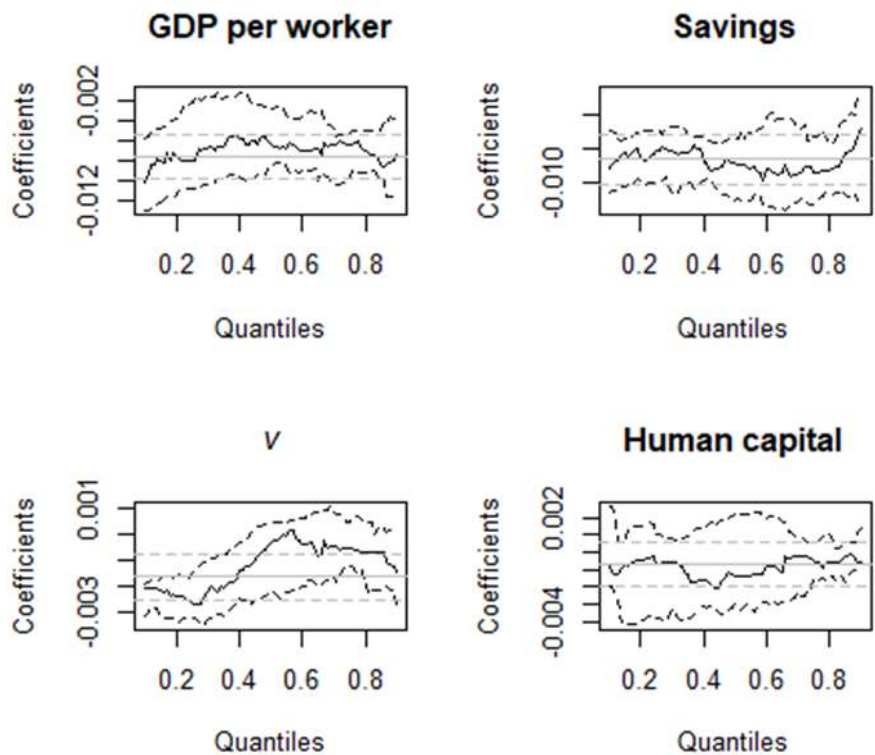

**Figure 1.** *Quantile regression effects (black line) of MRW covariates on GDP per worker growth. Confidence intervals at a level of 0.90 (black dotted lines). In grey OLS estimates and intervals (dotted).*

The output yields valuable information about $\beta$-convergence coefficients as we can observe different outcomes at each quantile level. In the tails, the effects of the initial level of GDP per-worker are lower so that the convergence speed is higher. The levels of the $\beta$ – coefficient around the median lie above the OLS $\beta$ coefficient of -0.0076, demonstrating that in the central part of the distribution the effect of the initial level of GDP is weaker. The presence of heterogeneous effects in the level of parameter estimates of the $\beta$ – coefficient provide some possibility of there being different dynamics among the regions.

Quantile regression approach have been used to evaluate economic performance (Kokic et al. 1997). Particularly, growth applications of quantile regression support an interpretation for which, at each level of the covariates, the units at higher levels of the conditional distribution outperform those at the lowest. In this application, this interpretation can be extended to understand how the higher (or lower) quantiles are influenced by MRW covariates, including the starting point level of GDP per worker. Despite the presence of these heterogeneous effects, for every quantile level $\beta$-convergence coefficients, are negative so that the hypothesis of conditional convergence for the 187 European regions cannot be rejected.

In the case of savings, the effect is more relevant at the lower quantile, while it is lower in the upper part of the conditional distribution. However, even if the magnitude of the estimates changes, the sign is negative at lower quantiles, and confirms previous studies on European convergence (Andreano et al. 2017). The sign of the savings coefficient is only positive in the extreme high tail, a situation that may relate to the ability of regions in the higher part of the conditional distribution to convert investments into economic growth; this result is consistent with other studies such as Barreto and Hughes (2004).

The effects of $v_i = \ln(n_i + l_i + d_i)$ are heterogeneous as they move from very low negative coefficients to positive ones. These differences highlight the negative effect of $v$ for units in the lower part of the conditional distribution, hence the negative effects of depreciation for those regions. The impact of $v$ increases particularly around the median in contrast to the OLS estimate.

The human capital coefficients do not show statistical significance. The magnitude tends to be homogeneous through the conditional distribution as it reveals no relevant changes apart from a small decrease in the median level. This provides evidence that the effect of human capital on GDP growth is weak; similar findings were presented in Canarella and Pollard (2004).

Standard quantile regression is able to estimate a family of linear functions and to generate a complete characterization of the conditional distribution, but the conventional model may be inappropriate in the case of significant endogeneity (Chernozhukov and Hansen 2006). Hence, the presence of spatial autocorrelation is likely to cause the quantile regression estimator to be inconsistent (Kim and Muller 2004).

Spatial quantile regression may be considered as a straightforward quantile generalization of the spatial lag model used in spatial econometrics. In our model, particularly, neighborhoods are built by using a connectivity matrix **W** defined as the $k$-nearest matrix accounting for the 5 nearest regions.[1]

Moreover, to estimate the model, we choose to use the IVQR estimator (Chernozhukov and Hansen 2005) with respect to the DSQR estimator to avoid problems connected to the use of *iid* data. Furthermore, the IVQR estimator is preferred to the ELQR estimator due to the additional computational costs of empirical likelihood.

The results of the spatial quantile regression are reported in figure 2 that highlights the parameters estimates for the four MRW covariates in the European regions. In the plot, a comparison is also offered between spatial quantile regression (in black) and standard quantile regression (in grey). Finally, confidence intervals are obtained using procedures adopted by Chernozhukov and Hansen (2006).

---

[1] Euclidean and other contiguity distance matrices were also considered, but the results obtained are not substantially different from those presented in the paper.

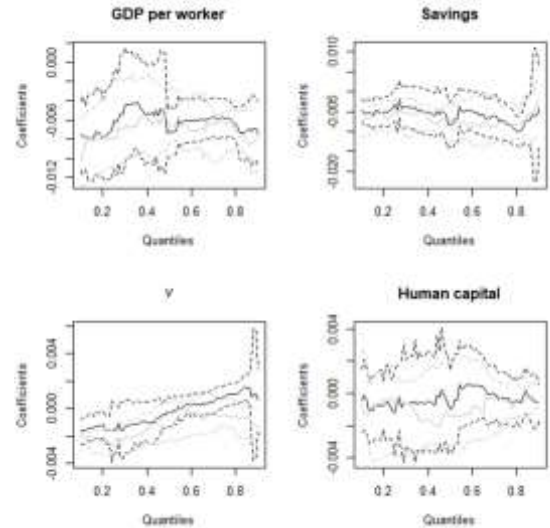

**Figure 2.** *Spatial quantile regression coefficients (black line) of MRW model. Confidence intervals at a level of 0.90 (black dotted lines), in grey standard quantile regression estimates parameters and intervals (dotted).*

Observing figure 2, we can note that larger differences between the two models appear particularly at the lower tail of the conditional distribution, where the $\beta$-convergence coefficients of spatial quantile regression are smaller than those observed in standard quantile regression.

In the case of savings, the results from the spatial quantile regression are similar to those obtained for the non-spatial specification. Major differences can be only identified in the central part of the distribution, where coefficients for the spatial specification are lower and moving toward zero. A slight increase of the savings coefficients in the extreme higher part is also present in the spatial specification equivalently to was observed for the non-spatial quantile regression.

For the variable $v$, the coefficients in the spatial model are lower in the tails and the curve for spatial quantile regression represents a situation in which the effect of $v$ is changing with the level of the quantile. This indicates that the effect of depreciation is reduced in magnitude by taking spatial effects into account. Around the median, no significant differences appear between the two quantile models.

Furthermore, the parameter estimates of human capital in the spatial model differ from the ones observed in the quantile regression, particularly in the central part of the conditional distribution. Nonetheless, the estimates of the parameters for the school attainment are largely insignificant as in the standard quantile model.

The spatial lag coefficient represents an important feature in a spatial model. Similarly, as for the other coefficients, the level of $\rho$ changes with the quantile. Spatial dependence in the form of spatial autocorrelation (see figure 3) is smaller in the lower part of the distribution, while it increases considerably in the upper tail. Note that confidence intervals are obtained according to Chernozhukov and Hansen (2006).

Spatial dependence is considerably important for the definition and interpretation of a regional model of economic growth. As our application involves data from European regions, much care must be taken when analyzing the magnitude of spatial autocorrelation in the model (see Kostov and Le Gallo 2015).

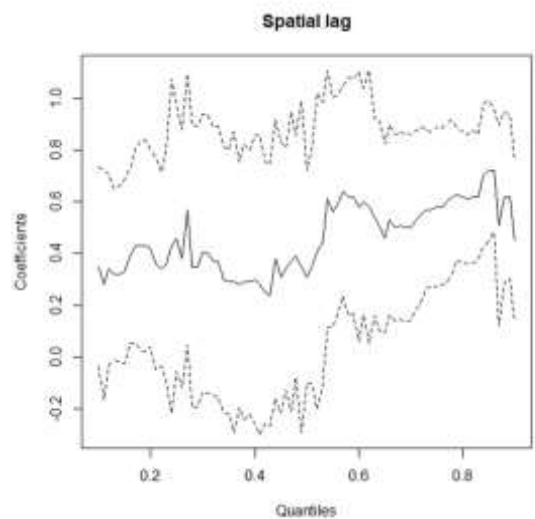

**Figure 3.** *Quantile estimates of spatial lag coefficients (black line). Confidence intervals at a level of 0.90.*

In the case of spatial quantile regression, this allows us to generate insights into the effect due to the average growth of a neighborhood at a certain quantile (Liao and Wang 2012). Hence, the presence of heterogeneous responses to nearby levels of growth is considered.

Since the spatial literature has been focused on considering the expected value of the effects of spatial interactions, adopting a quantile framework may help to point to the underlying strength of spatial interconnections conditioned to the level of quantile.

The presence of heterogeneous effects of neighborhoods requires some caution in the interpretation of the spatial lag models. In our specific case, the presence of stronger interactions at higher levels of the conditional distribution suggests that greater attention needs to be paid to the presence of heterogeneous economic dynamics.

Moreover, in the European regions, the variations of the $\rho$ coefficient suggest that the effects of spatial interactions are increasing at the level of quantile. The spatial lag parameter is significant in the upper part, a situation that is strictly connected to the increase in representativeness. Problems of misspecification are better solved by the presence of a spatial parameter in the upper part of the conditional distribution, where spatial coefficients tend to be more significant.

In the last 20 years, a large literature (Liao et al. 2006; Postiglione et al. 2013) emphasized the presence of both types of spatial effects, spatial dependence and spatial heterogeneity, in economic convergence of European regions. Specifically, the presence of spatial dependence has been considered as a significant issue for the estimation of an economic growth model (LeSage and Fischer 2008). Conversely, spatial heterogeneity is likely to cause weak global convergence and the necessity to consider the presence of multiple regimes based on the differences in the economic motion across regions. Identifying clusters of regions is a foremost concern in the analysis of regional economic convergence.

In figure 4, a map of the clusters of EU NUTS 2 regions is presented. In the map, regions are classified according three different residual intervals to measure the gap from the best achievement in terms of conditional growth. Particularly, a quantile model at $\tau = 0.90$ is selected as the best achievement similarly to efficiency analysis. Hence, choice of the number of groups, $k$, is made in an exogenous way, comparing the maps for different $k$'s. In this case, $k=3$ is chosen to clearly visualize differences in the groups.

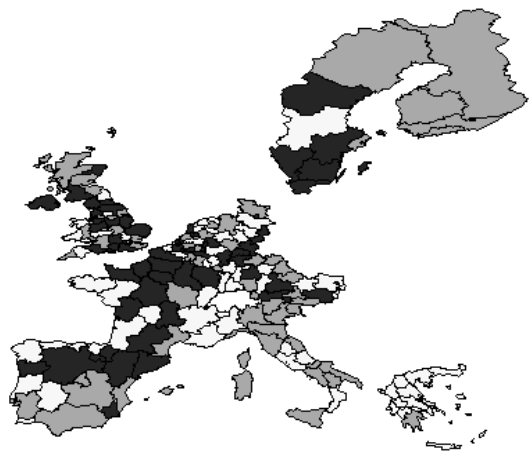

**Figure 4.** *Map of residuals for the quantile model for $\tau=0.90$. Black indicates lower levels of conditional distribution, grey the median interval, white indicates class closer to the estimated quantile.*

Figure 4 represents a scenario in which a number of regions of Central and Southern Europe are situated near the upper quantile justifying the higher level of spatial autocorrelation in the higher part of the conditional distribution (see figure 4, regions in white). Additionally, regions in the West of France and the region of Svealand (Sweden) are situated near the upper quantile. Regions clustered in the lower part of the conditional distribution tend to be more scattered across the North of Europe and the North of Spain, including mainly regions of Sweden, UK, North of France and Paris area, North of Spain and West Germany (see figure 4, regions in black).

# 6. Conclusion

In this paper, we analyzed the conditional $\beta$-convergence of 187 European NUTS 2 regions by adopting an MRW specification of the economic growth function. The use of quantile regression allows us to investigate the presence of heterogeneous effects depending on the quantiles. The signs of the $\beta$-coefficients across the conditional distribution support that convergence is a confirmed hypothesis. However, the magnitude of $\beta$-convergence varies across the quantiles so that lower quantiles show faster speed of convergence with respect to the higher ones.

Further, we also adopted a spatial specification of quantile regression considering spatial dependence between European regions. The results from the spatial quantile approach show that problems of misspecification may affect non-spatial quantile regression when considering spatial units. Particularly, by using spatial quantile regression we appreciate a considerable increase in the fit and differences in the estimates of parameters, particularly for the $\beta$-coefficients. Moreover, the focus on heterogeneous effects includes also the spatial lag parameter that assumes different values across quantiles.

The presence of heterogeneous effects in all the covariates of the MRW model indicate that economic growth still is a heterogeneous process stressing the importance of a major understanding of causes of disparities. However, we aware that this method is only a first step in an expanded use of quantile regression for the analysis and the modeling of spatial heterogeneity.